\documentclass[twocolumn,twoside,slac_two]{revtex4}
\usepackage{graphicx}
\usepackage{fancyhdr}
\usepackage{hyperref}
\usepackage{float}

\hypersetup{
    colorlinks=true,
    urlcolor=magenta,
}
\begin{document}

\title{DAMPE space mission: first data}


\author{F.Gargano, on behalf of DAMPE Collaboration}
\affiliation{INFN Bari, via Orabona 4, 70125 Bari, Italy}

\begin{abstract}
The DAMPE (DArk Matter Particle Explorer) satellite was launched on December 17, 2015 and started its data taking operation a few days later.

DAMPE has a  large geometric factor ($\sim~0.3\ m^2\ sr$) and provides good 
tracking, calorimetric and charge measurements for electrons, gammas rays and nuclei. This will 
allow precise measurement of cosmic ray spectra from tens of $GeV$ up to about $100\ TeV$. In particular, 
the energy region between $1-100\ TeV$ will be explored with higher 
precision compared to previous experiments. 
The various subdetectors allow an efficient identification of the electron signal 
over the large (mainly proton-induced) background. As a result, the all-electron spectrum will 
be measured with excellent resolution from few $GeV$ up to few $TeV$, thus giving the opportunity
to identify possible contribution of nearby sources. A report on the mission goals and status is presented, together with the on-orbit detector performance and the first data coming from space.
\end{abstract}

\maketitle

\thispagestyle{fancy}


\section{Introduction} \label{sec:intro}

The DArk Matter Particle Explorer (DAMPE) is a space mission supported by the strategic space projects of
the Chinese Academy of Sciences with the contribution of Swiss and Italian institutions~\cite{Chang2014,ambrosi}. 
The rocket has been successfully launched on December 17, 2015 %(Fig.~\ref{fig:launch}) 
and DAMPE presently flies regularly 
on a sun-synchronous orbit at the altitude of $500\ km$. The satellite is equipped with four different detectors: 
a plastic scintillator array, a silicon-tungsten tracker, a BGO calorimeter and a neutron detector. They are 
devoted to measure the fluxes of charged CRs (electrons, protons and heavier nuclei), to study the high energy 
gamma ray signal from astrophysical sources and to search for indirect dark-matter signatures.

%\begin{figure}[ht]
%\centerline{\includegraphics[width=8.5cm,height=6.0cm]{Images/launch.pdf} }
%\vspace {-0.0cm}
%\caption{Jiunquan base in China (December 17, 2015). The launch of the rocket with the DAMPE 
%satellite~\cite{ambrosi}.}
%\label{fig:launch} 
%\end{figure}

%\vspace{-3.6cm}
\section{On-board instruments} \label{sec:instru}
%\vspace{-3.4cm}

\begin{figure}[h]
\centerline{\includegraphics[width=8.5cm,height=5.4cm]{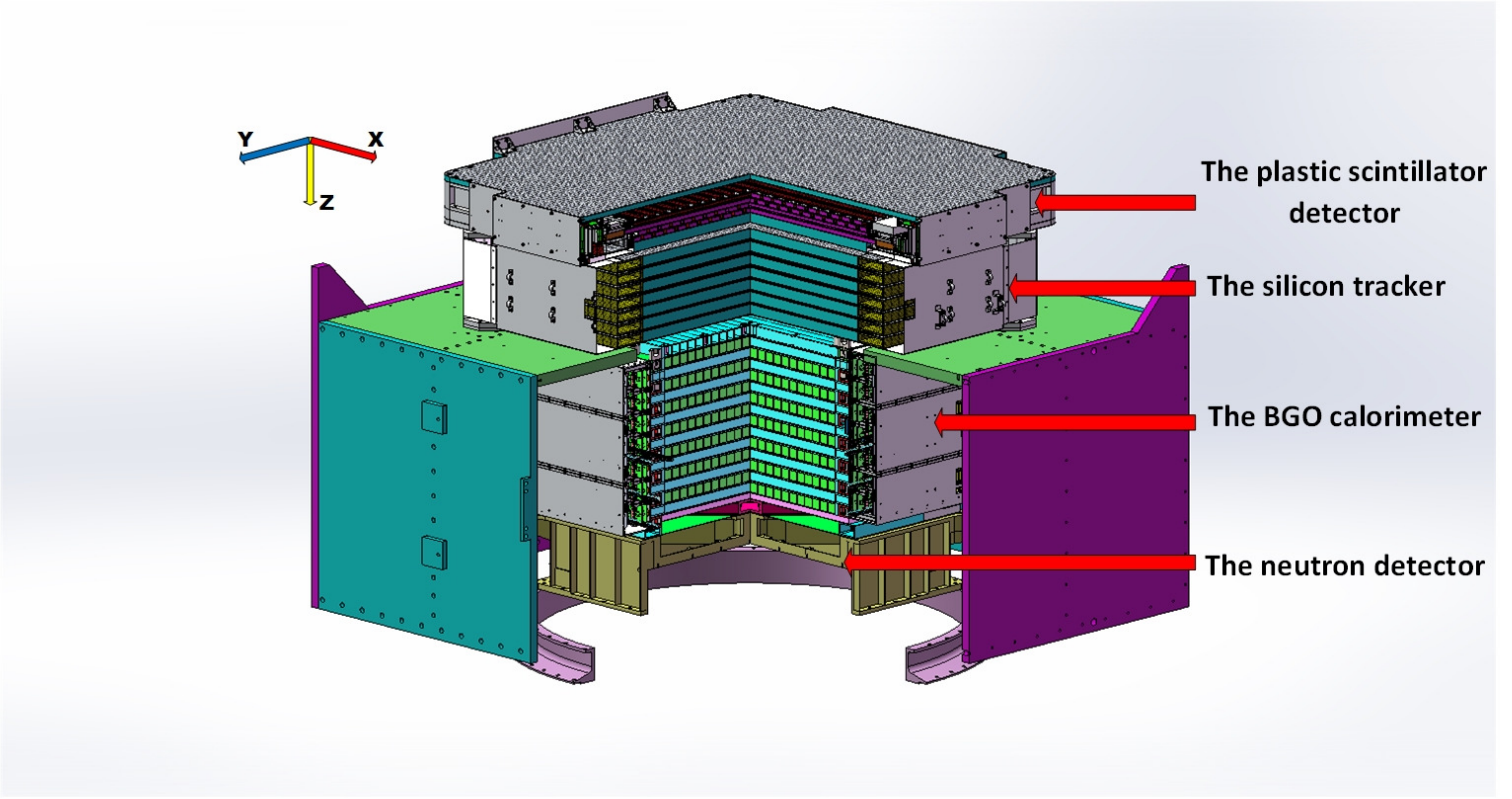} }
\vspace {-0.0cm}
\caption{Side view of the DAMPE detector~\cite{ambrosi}.} \label{fig:side} 
\end{figure}

DAMPE (Fig.~\ref{fig:side}) consists of a Plastic Scintillator strip Detector (PSD) that is used as anti-coincidence 
and charge detector, a Silicon-Tungsten tracKer-converter (STK) to reconstruct the direction of incident particles, 
a BGO imaging calorimeter (BGO) of about 32 radiation lengths that measures the energy with high resolution 
and distinguishes between electrons and protons, and a NeUtron Detector (NUD) that can further increase the hadronic 
shower rejection power.

%\vspace{-3.4cm}
\subsection{Plastic Scintillator Detector (PSD)}  \label{sec:PSD}
%\vspace{-3.4cm}

The high energy sky is mainly dominated by nuclei with different electrical charges~\cite{PSD2016} . This charged flux is
studied by DAMPE but it is also a background for gamma astronomy. Therefore the PSD is 
designed to work as a veto and to measure the charge ($Z$) of incident high-energy particles up to $Z = 26$.
Following these requirements the PSD must have a high detection efficiency for charged particles, a large 
dynamic range and a relatively good energy resolution.
\begin{figure}[h]
\centerline{\includegraphics[width=5.95cm,height=3.78cm]{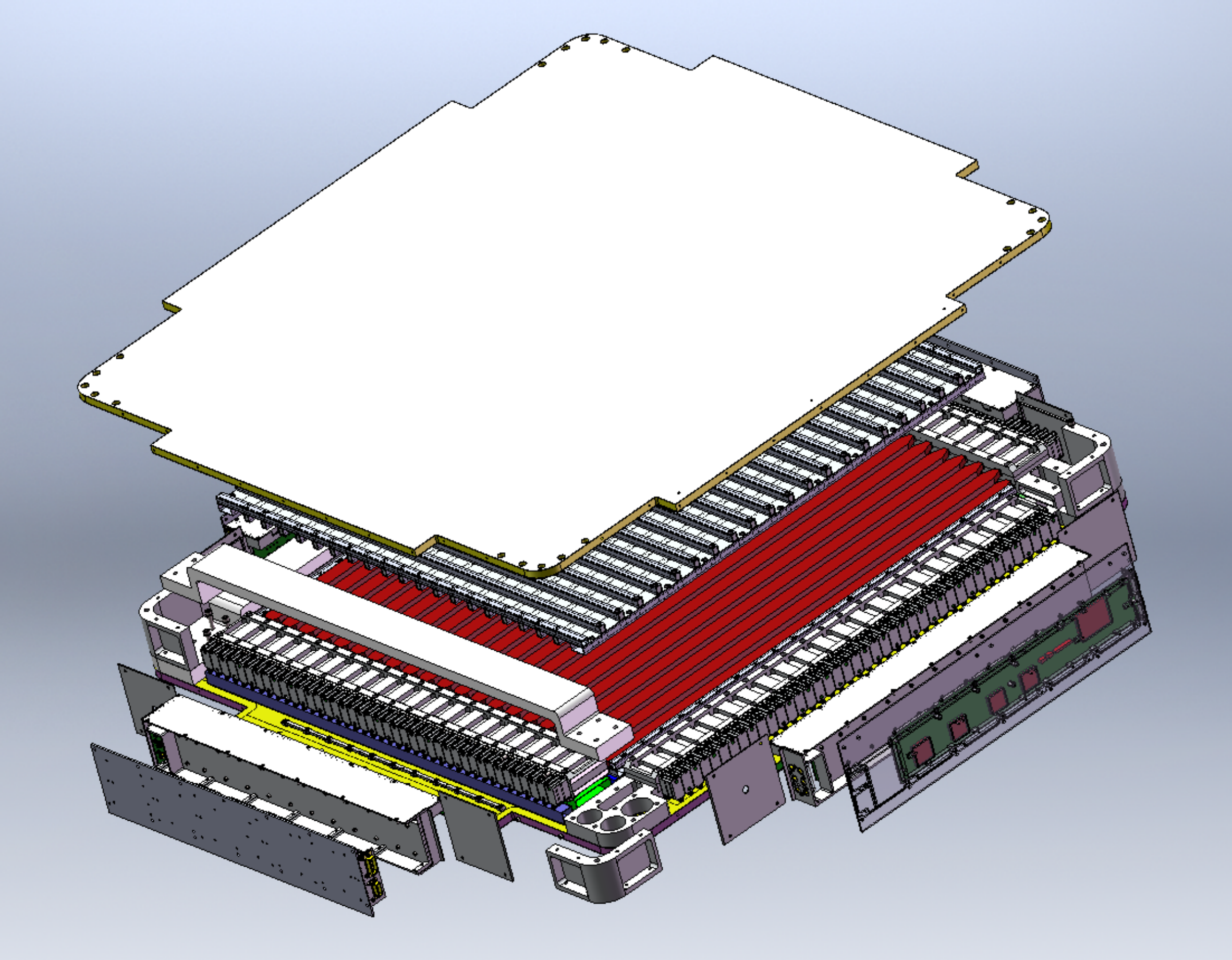} }
\vspace {-0.0cm}
\caption{Exploded view of the PSD~\cite{ambrosi}.} \label{fig:PSD}	
\end{figure}

\subsection{Silicon-Tungsten Tracker (STK)} \label{sec:STK}
%\vspace {+0.2cm}

\begin{figure}[h]
\centerline{\includegraphics[width=8.5cm,height=5.4cm]{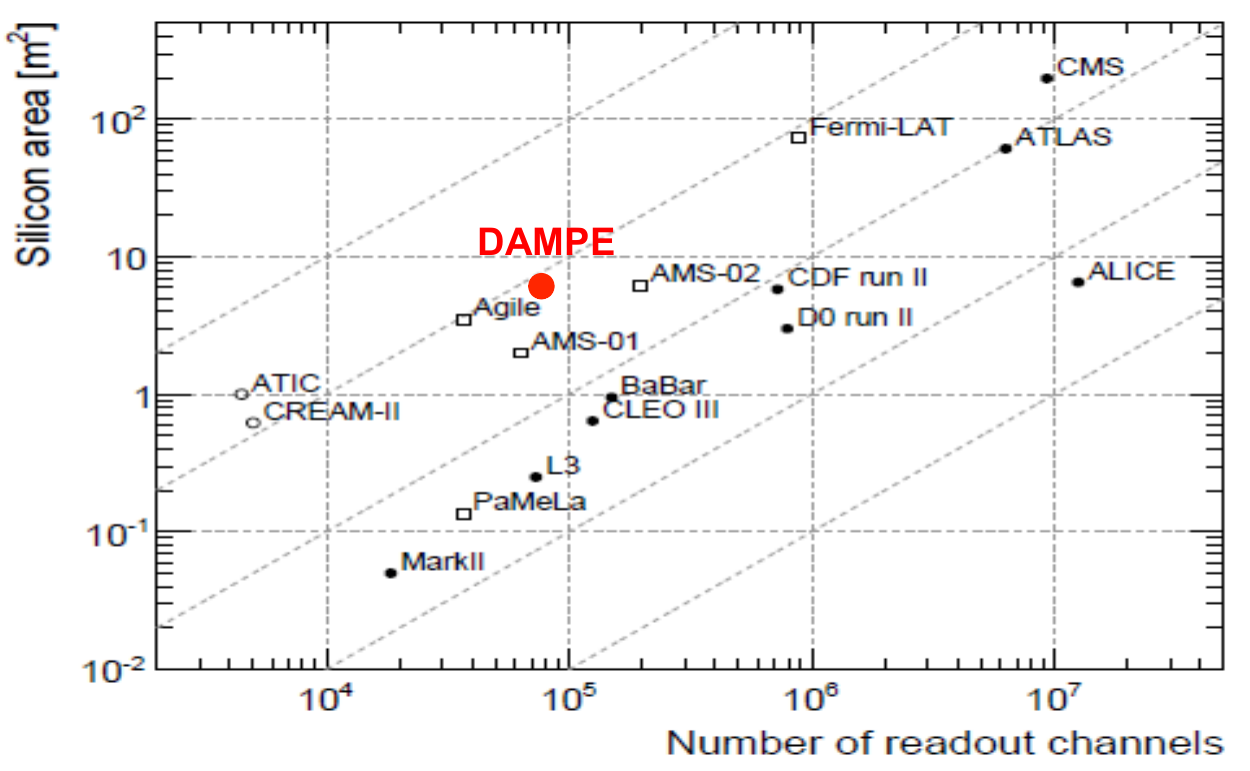} }
\vspace {-0.0cm}
\caption{Comparison of STK features with other detectors.} \label{fig:comparison} 
\end{figure}

The Silicon-Tungsten tracKer-converter 
%(Fig.~\ref{fig:stk}) 
is devoted to the precise reconstruction of the particle
tracks. It consist of twelve position-sensitive silicon detector planes (six planes for the $x$-coordinate, six planes 
for the $y$-coordinate). Three layers of tungsten are inserted in between the silicon planes (2, 3, 4 and 5) to 
convert gamma rays in electron-positron pairs. The specifications of the STK are given in Table~\ref{tab:STK} and a comparison with other 
experiments is shown in Fig.~\ref{fig:comparison} for what concerns the active area and the number of channels.

\begin{table} \begin{center}
\caption{STK specifications} \label{tab:STK}
\begin{tabular}{lll} \hline
Active area of each silicon layer & $0.5534\ m^2$ \\
Silicon thickness   		  & $320\ \mu m$  \\
Silicon strip pitch		  & $121\ \mu m$  \\
Tungsten thickness       	  & $1\ mm$       \\
Fraction of radiation length      & 0.976         \\
Power consumption		  & $82.7\ watts$ \\
Mass				  & $154.8\ kg$   \\ \hline
\end{tabular} \end{center} \end{table}

\subsection{BGO Calorimeter (BGO)} \label{sec:BGO}

\begin{figure}[h]
\vspace{-0.4cm}
\centerline{\includegraphics[width=8.5cm,height=5.4cm]{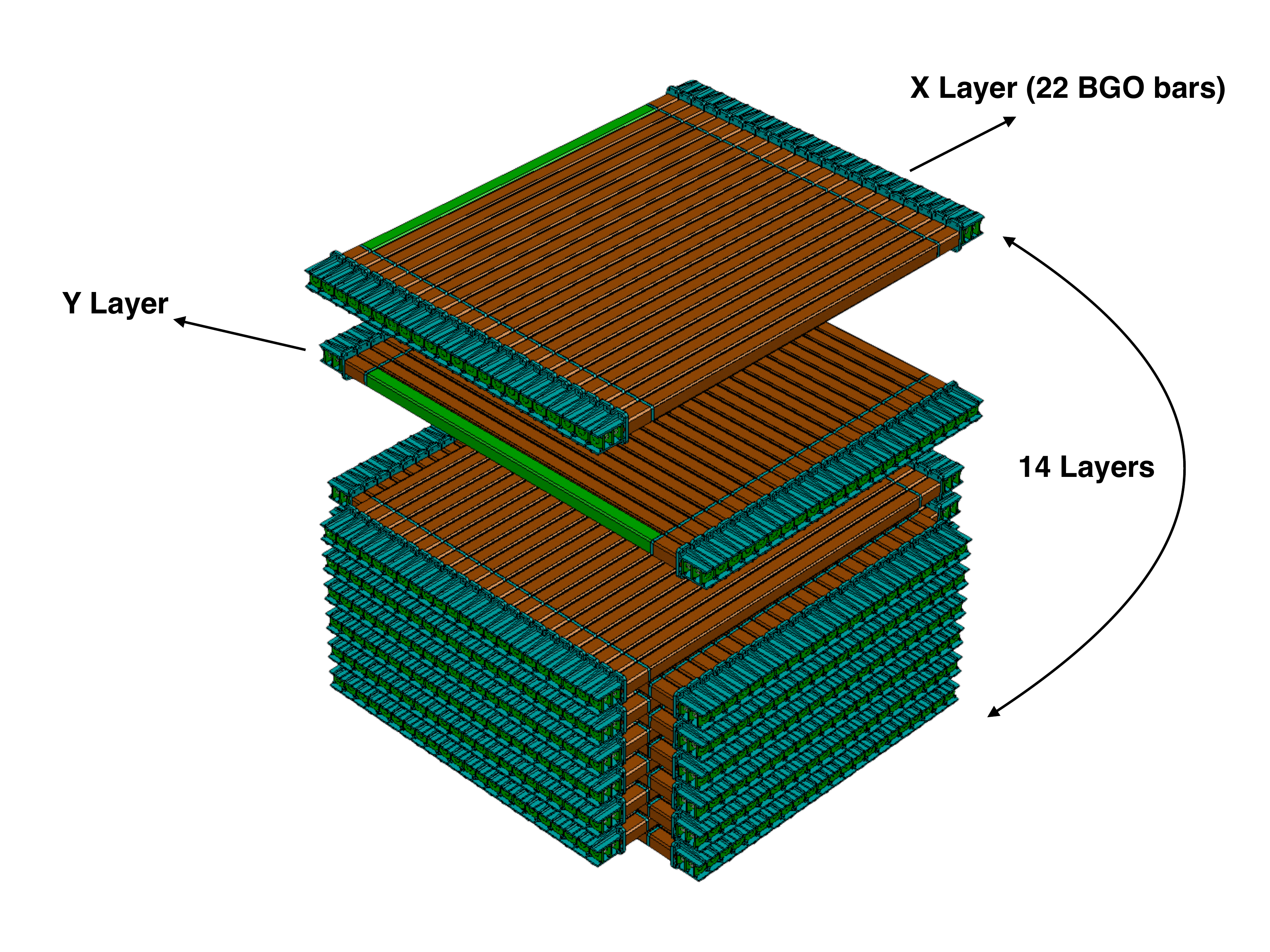} }
\vspace {-0.0cm}
\caption{Exploded view of the BGO scintllator~\cite{ambrosi}.} \label{fig:bgo} 
\end{figure}

The BGO calorimeter is used to measure the energy deposition of incident particles and to reconstruct the 
shower profile ~\cite{BGO2012}. The trigger of the whole DAMPE system is based on the signals from the BGO. The reconstructed shower profile is fundamental to distinguish between electromagnetic and hadronic showers.\\
\indent The calorimeter is composed of 308 BGO crystal bars ($2.5 \times 2.5 \times 60\ cm^3$ is the volume
of a single bar). The crystals are optically isolated from each other and are arranged horizontally in 14 
layers of 22 bars (Fig.~\ref{fig:bgo}). The bars of a layer are orthogonal to those of the adjacent plane
in order to reconstruct the shower in both views ($x-z$ and $y-z$).
The total vertical depth of the calorimeter is about 32 radiation lengths and 1.6 nuclear interaction lengths. 
%This unprecedented depth ensures that almost $100\%$ of the energy of electrons and $\gamma$-rays is deposited in 
%the calorimeter, and about $40\%$ of proton energy. The high limit ($\sim 100\ TeV$) of the DAMPE measurement 
%range is essentially determined by the overall geometric factor and the calorimeter dynamic range.
Table~\ref{tab:bgo} summarizes the key parameters of BGO calorimeter.

\begin{table} \begin{center}
\vspace{-0.3cm}
\caption{BGO specifications} \label{tab:bgo}
\begin{tabular}{llllll} \hline
Active area	     	  & $60 \times 60\ cm^2$ (on-axis) \\
Depth                     & 32 radiation lengths           \\
Sampling     		  & $> 90\%$                       \\
Longitudinal segmentation & 14 layers                      \\
Lateral segmentation      & $\sim 1$ Moliere radius        \\ \hline
\end{tabular} \end{center} \end{table}

%%%%%%%%%%%%%%%%%%%%%%
%\clearpage\newpage
%%%%%%%%%%%%%%%%%%%%%%

\subsection{Neutron Detector (NUD)} \label{sec:NUD}

The NeUtron Detectos is a further device to distinguish the 
types of high-energy showers. It consist of four boron-loaded plastics each read out by a PMT.
Typically hadron-induced showers produce roughly one order of magnitude more neutrons than electron-induced showers.
Once these neutrons are created, they thermalize quickly in the BGO calorimeter and the neutron activity can be 
detected by the NUD within few $\mu s$ ($\sim~2\ \mu s$ after the shower in BGO).  
Neutrons entering the boron-loaded scintillator undergo the capture process
$$^{10}B + n \rightarrow ^7Li + \alpha + \gamma$$
Its probability is inversely proportional to neutron velocity and the capture time is inversely proportional to $^{10} B$ loading. 
Roughly 570 optical photons are produced in each capture~\cite{Drake1986}. %\\

\section{Detector design and ground tests} \label{sec:BT_CERN}

\begin{figure}[h]
\centerline{\includegraphics[width=8.0cm,height=5.4cm]{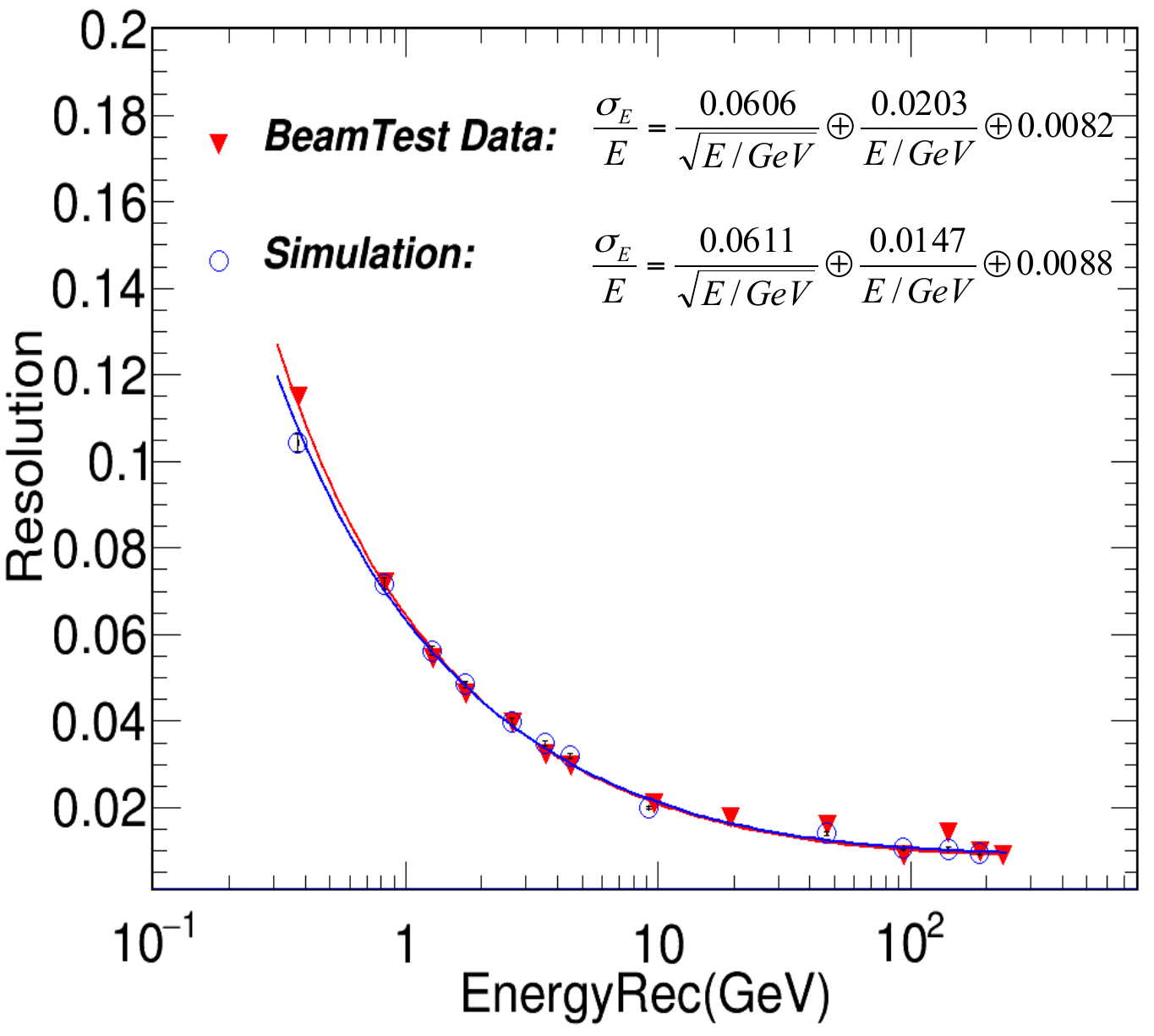} }
\vspace {-0.0cm}
\caption{Energy resolution for electromagnetic showers - Preliminary comparison of beam test data with simulation ~\cite{ZhangZY2016}.} \label{fig:resol} 
\end{figure}

An extensive Monte Carlo simulation activity was carried out during the R\&D phase  in order to find a proper compromise between
research goals and limitations on geometry, power consumption and weight. DAMPE performances were verified by a series 
of beam tests at CERN. %(Fig.~\ref{fig:cern}). 
The PS and SPS accelerators provide electron and proton beams. The beam test data were used to study the performance of the BGO calorimeter, and in particular the energy resolution (Fig.~\ref{fig:resol}), the linearity and the  $e/p$ separation. Also a beam of argon fragments was used for performing tests with heavy ions. 
Details of the beam-test preliminary results as well as the features of the qualified module can be found 
in~\cite{Chang2014,Gallo2015,Azzarello2016,ZhangZY2016}.%\\

\vspace {-0.2cm}
\section{On-orbit operation} \label{sec:on-orbit}

After launch, the spacecraft entered the sky-survey mode immediately and the dedicated-calibration of the detector was 
performed in two weeks. The calibration included the studies of pedestal, response to MIPs, alignment, timing, etc... %Comparing  

\indent The satellite is on a solar-synchronized orbit lasting 95 minutes. The pedestal calibration is performed twice per orbit and
the global trigger rate is kept at $\sim 70\ Hz$ by using different pre-scales for unbiased and low-energy triggers at different 
latitudes. 
In  absence of on-board analysis processing, the data are just packaged with timestamp 
and transmitted to ground (about 4 millions of events per day corresponding to $15\ GB$). After the event reconstruction the
data size is $100\ GB$ per day. 

\section{First on-orbit data and performances} \label{sec:data}

The DAMPE detectors started to take physics data very soon after the launch. The performance parameters
(temperature, noise, spatial resolution, efficiency) are very stable and very close to what expected.
The absolute calorimeter energy measurement has been checked by using the geomagnetic cut-off and it results well
calibrated. Also the absolute pointing has been successfully verified. The photon-data collected 
in 165 days were enough to draw a preliminary high-energy sky-map where the main gamma-ray sources are visible 
in the proper positions.

\begin{figure}[h]
\centerline{\includegraphics[width=8.5cm,height=5.4cm]{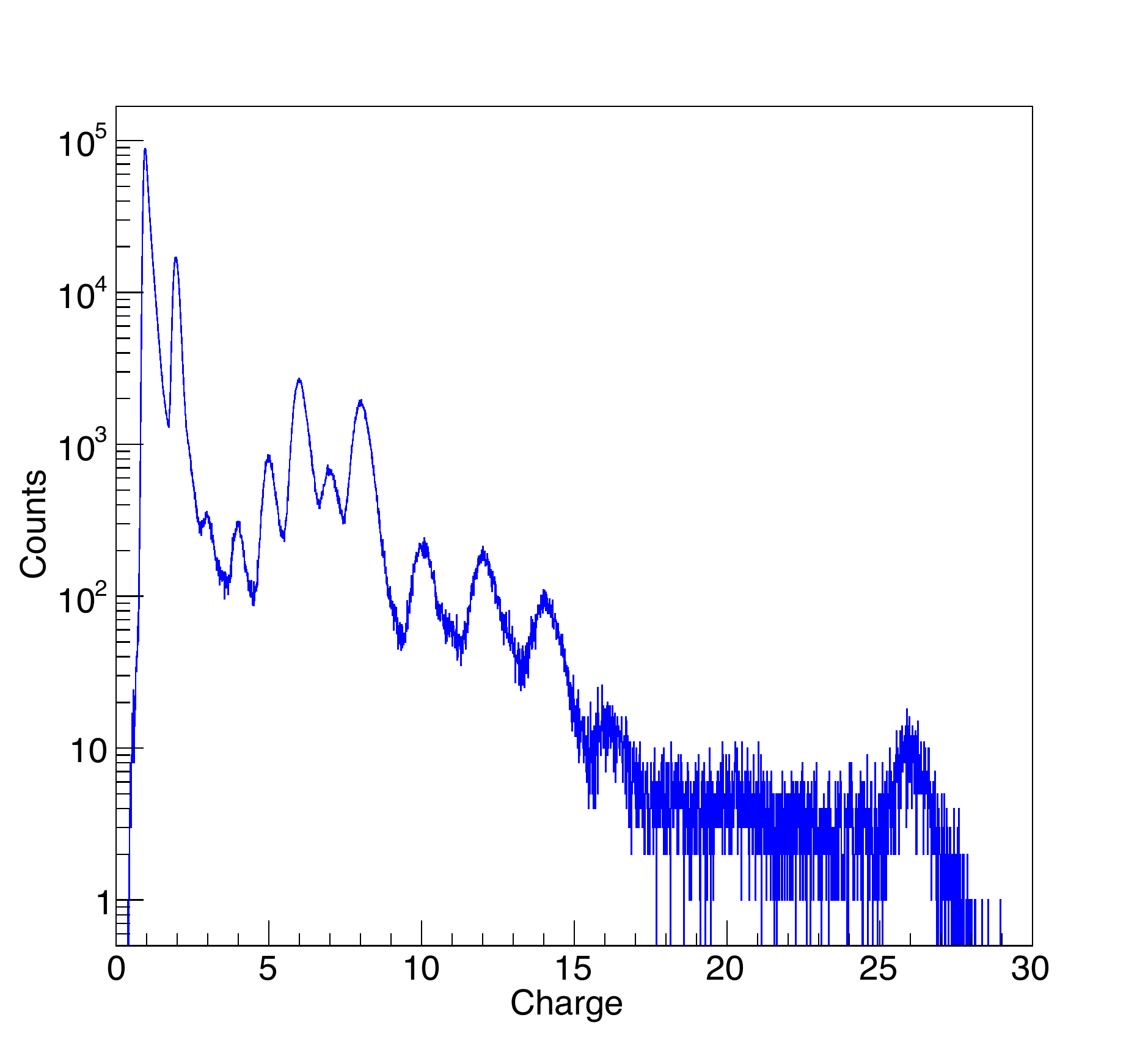} } %% Charge_10days.pdf} }
\vspace {-0.0cm}
\caption{Very preliminary $Z$ measurement up to iron with only 10 days of data.} \label{fig:z2} 
\end{figure}

\indent The energy released in the PSD allows to measure the charge and to distinguish the different 
nuclei in the CR flux. %Figs.~\ref{fig:z1} and 
Fig. \ref{fig:z2} show the result of this measurement
for the full range up to iron ($1 \le Z \le 26$).

\begin{figure}[h]
\centerline{\includegraphics[width=8.0cm,height=5.4cm]{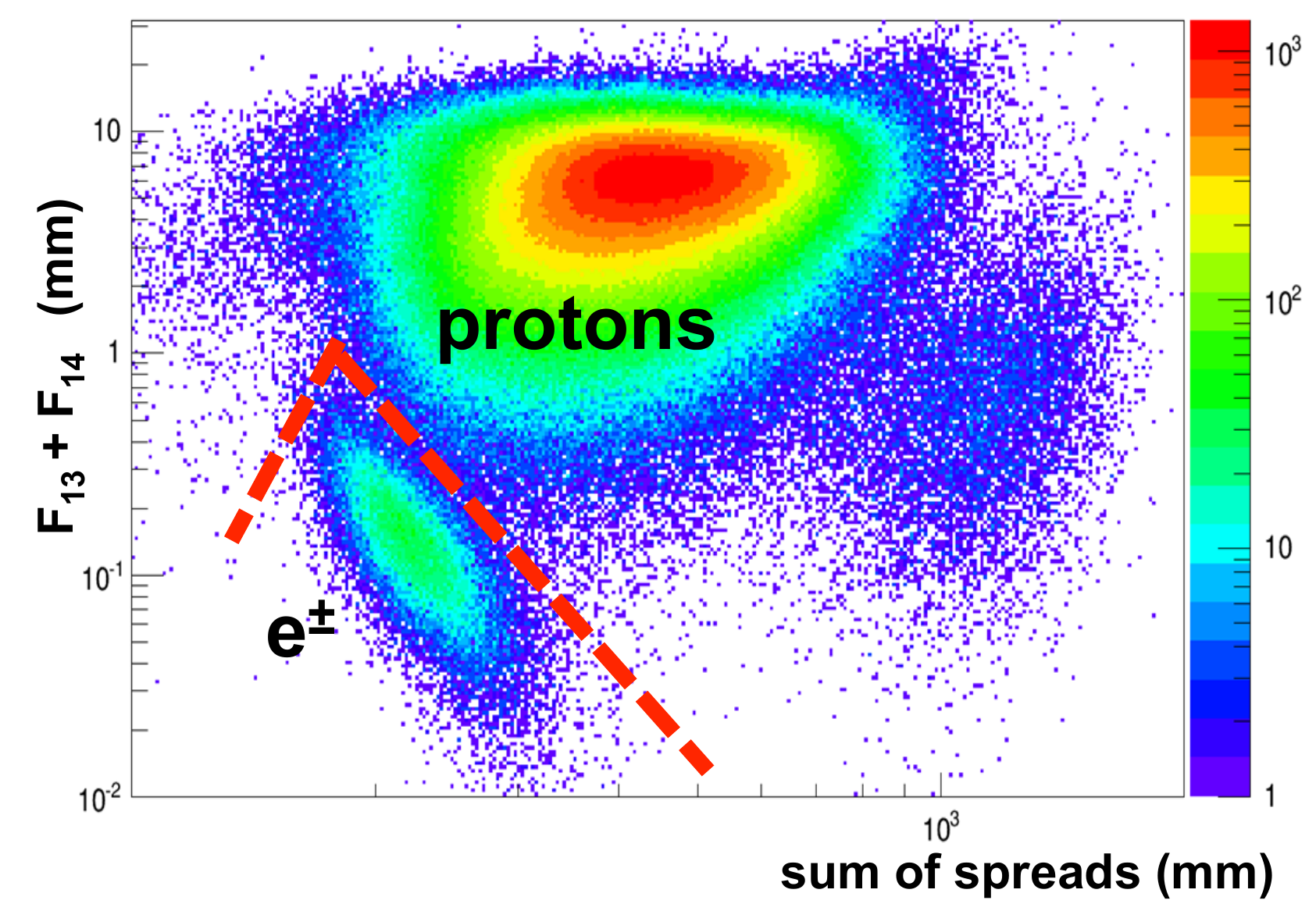} }
\vspace {-0.0cm}
\caption{Preliminary result about e/p separation. 
%% The lepton sample is well separated from hadron sample. 
The background on the right is due to CRs entering the satellite from the sides.} \label{fig:fig7} 
\end{figure}

The measurement of electron and positron flux is one of the main goals of the DAMPE mission because 
some dark matter signature could be found in the electron and positron spectra. The shower development in the BGO provides
the main tool to distinguish leptons from hadrons. Then a shape parameter is defined as:
$$F_i = spread_i \times \frac{E_i}{E_{tot}},$$
where $i$ is the index of the BGO layer ($1 \le i \le 14$), $spread_i$ is the shower width in the i-th layer, $E_i$ and $E_{tot}$ are 
the energy on the single layer and on all the layers, respectively. Using the shape parameters on the last BGO layers
(13, 14) it is possible to separate leptons from hadrons with a rejection power higher than $10^5$ (preliminary result
in Fig.~\ref{fig:fig7}). The rejection capability will be further enhanced by means of the NUD.

\section{Expected measurements in 3 years} \label{sec:what}

\begin{figure}[h]
\centerline{\includegraphics[width=8.0cm,height=5.4cm]{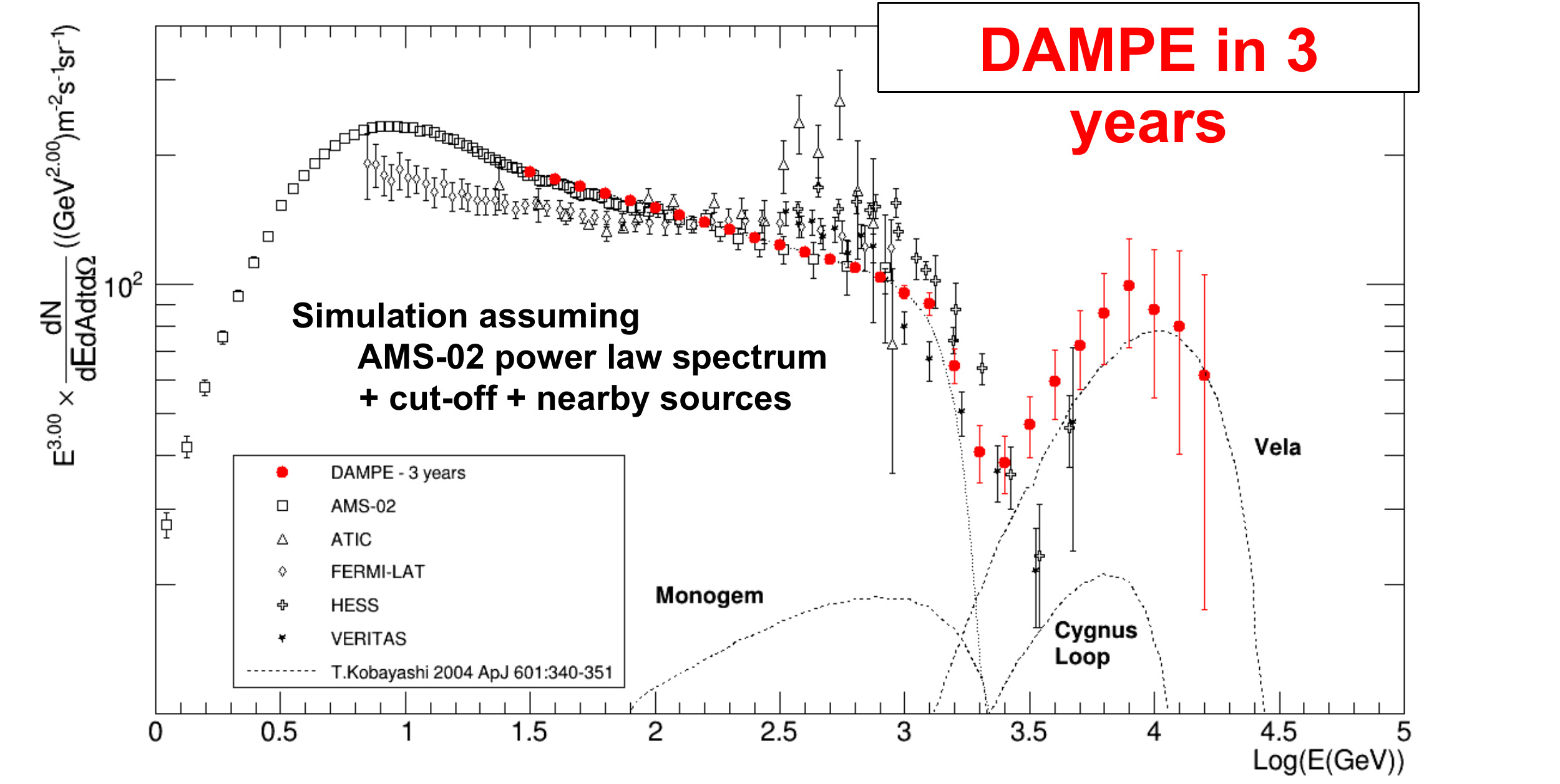} }
\vspace {-0.0cm}
\caption{All-electron spectrum. The red dots represent the possible DAMPE measurements
in 3 years assuming the power law suggested by the AMS-02 experiment, a cut-off at
$\sim~1\ TeV$ and nearby astrophysical sources.} \label{fig:fig8} 
\end{figure}

The DAMPE detector is expected to work for more than 3 years. This data-taking time is sufficient 
to investigate deeply many open questions in CR studies. In Fig.~\ref{fig:fig8} the possible 
DAMPE measurement of the all electron spectrum in 3 years is shown. The energy 
range is so large to observe a cut-off and a new increase of the flux due to nearby astrophysical 
sources, if present.

\begin{figure}[H]
%% \vspace {-6.5cm}
\centerline{\includegraphics[width=8.0cm,height=5.4cm]{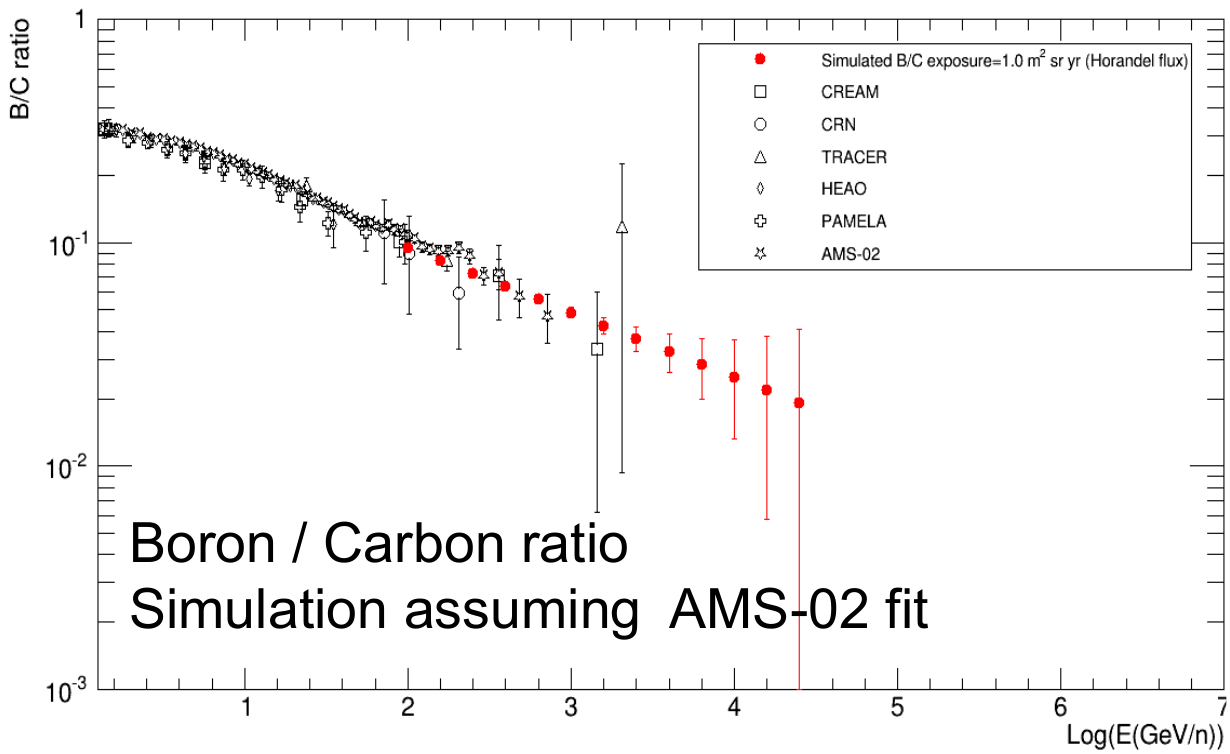} }
\vspace {-0.0cm}
\caption{Boron/carbon ratio versus energy per nucleon. The red dots 
indicate the possible DAMPE measurement in 3 years assuming the AMS-02 data fit.} \label{fig:fig11} 
\end{figure}

\indent Many experiments~\cite{atic,cream,pamela,ams} observed a hardening of the CR elemental
spectra at $TeV$-energies. This is another interesting topic related to CR origin and propagation and DAMPE 
will be able to perform significant measurements about it 
%(Fig.s~\ref{fig:fig9} and ~\ref{fig:fig10}) 
and also about the boron/carbon ratio (Fig.~\ref{fig:fig11}). Finally the large exposure will allow
extending energy spectra measurements for protons and nuclei up to tens of $TeV$.

\section{Conclusions} \label{sec:conclu}

The DAMPE satellite has been successfully launched in orbit on December 2015 and the preliminary data 
analyses confirm that the detectors work very well. The DAMPE program foresees important measurements 
on the CR flux and chemical composition, electron and diffuse gamma-ray spectra and anisotropies, gamma 
astronomy and possible dark matter signatures. This challenging program is based on the outstanding 
DAMPE features: the large acceptance ($0.3\ m^2\ sr$), the "deep" calorimeter ($32\ X_0$), the precise 
tracking and the redundant measurement techniques.

\bigskip % extra skip inserted

\end{document}